# An alternative Interpretation of residual feed intake by phenotypic recursive relationships in dairy cattle

Xiao-Lin Wu,[1,2]* Kristen L. Parker Gaddis,[1] Javier Burchard,[1] H. Duane Norman,[1] Ezequiel Nicolazzi,[1] Erin E. Connor,[3] John B. Cole,[4] and Joao Durr[1]

[1] Council on Dairy Cattle Breeding, Bowie, MD 20716
[2] Department of Animal and Dairy Sciences, University of Wisconsin, Madison, WI 53706
[3] Department of Animal and Food Sciences, University of Delaware, Newark, DE 19716
[4] USDA, Agricultural Research Service, Animal Genomics and Improvement Laboratory, Beltsville, MD 20705-2350

* Corresponding author: nick.wu@uscdcb.com

**ABSTRACT**

There has been an increasing interest in residual feed intake (RFI) as a measure of net feed efficiency in dairy cattle. RFI phenotypes are obtained as residuals from linear regression encompassing relevant factors (i.e., energy sinks) to account for body tissue mobilization. However, fitting energy sink phenotypes as regression variables in standard linear regression was criticized because phenotypes are subject to measurement errors. Multiple-trait models have been proposed which derive RFI by follow-up partial regression. By re-arranging the single-trait linear regression, we showed a causal RFI interpretation underlying the linear regression for RFI. It postulates recursive effects in energy allocation from energy sinks on dry matter intake, but the feedback or simultaneous effects are assumed to be nonexistent. A Bayesian recursive structural equation model was proposed for directly predicting RFI and energy sinks and estimating relevant genetic parameters simultaneously. A simplified Markov chain Monte Carlo algorithm that implemented the Bayesian recursive model was described. The recursive model is asymptotically equivalent to one-step linear regression for RFI, yet extends the analytical capacity to multiple-trait analysis. It is equivalent to Bayesian-implemented, multiple-trait model reparameterized based on Cholesky decomposition of phenotypic (co)variance matrix for evaluating RFI, but varied in assumptions about relationships between energy sinks.

## Technical note

In the past decades, residual feed intake (RFI) has become increasingly popular to measure net feed efficiency. RFI was initially proposed by Koch et al. (1963) as the residuals from linear regression (LR) of feed intake on various energy sinks. It represents a resource allocation theory, which partitions feed intake into the feed intake expected for the given production level and a residual portion (Herd, 2009). Genetic evaluation on RFI often takes two stages (Berry and Crowley, 2013). In the first stage, dry matter intake (DMI) is taken to be a linear function of variables (i.e., energy sinks) to account for body tissue mobilization. In dairy cattle, energy sinks often include metabolic body weight (MBW), energy-corrected milk (ECM) or milk net energy (MILKNE), body condition score, and changes in body weight (ΔBW) (e.g., Templeman et al., 2015; Pryce et al., 2019; Løvendahl et al., 2018; Islam et al., 2019). In the second stage model, the computed RFI phenotypes are fitted by a mixed-effects model to estimate RFI genetic values and relevant genetic parameters. Combining these two modeling stages leads to one-step LR for RFI (e.g., Templeman et al., 2015; Løvendahl et al., 2018), eliminating the need to estimate the residuals as the RFI phenotypes specifically. Fitting phenotypes as regressor variables in LR was criticized (Lu et al., 2015) because standard regression models assume that regressor variables have been measured precisely. In reality, however, phenotypes are subject to measurement errors. Multiple-trait models have been proposed which obtain RFI genetic values indirectly through follow-up partial regression (Kennedy et al., 1993; Lu et al., 2015; Islam et al., 2020; Tempelman and Lu, 2020). Dependent variables include DMI and energy sinks, but noting that some researchers (e.g., Lu et al., 2015) recommended treating ΔBW as a covariate because its heritability was low. With various methods available, their biological implications for the computed RFI remain to be exploited (Martin et al., 2020). By re-arranging the LR equation, we came across an alternative, causal interpretation of RFI by phenotype recursiveness between DMI and energy sinks.

Consider a single animal, say $i$. Let $y_{i1}$ be a variable for DMI phenotypes and let $y_{i2}$, …, and $y_{ik}$ be variables representing the phenotypes for $k$-$1$ energy sinks, all measured on this animal. A simple energy model includes only energy sinks as fixed effects, plus the residual ($r_i$) as a RFI phenotype (Løvendahl et al., 2018):

$$y_{i1} = \sum_{j=2}^{k} \lambda_{1j} y_{ij} + r_i \tag{1}$$

where $\lambda_{1j}$ quantifies the effect of energy sink $j$ on DMI. The energy sink model may include additional covariates or factors (Templeman et al., 2015; Pryce et al., 2015). Then, the residual is fitted by a mixed-effects model:

$$\hat{r}_i = \mu_1 + \mathbf{x}'_{i1}\boldsymbol{\beta}_1 + \mathbf{z}'_{i1}\boldsymbol{a}_1 + e_{i1} \tag{2}$$

where $\mu_1$ is the overall mean, $\boldsymbol{\beta}_1$ is a vector of fixed effects, $\boldsymbol{a}_1$ is a vector of random animal additive genetic values, $\mathbf{x}_{i1}$ and $\mathbf{z}_{i1}$ are the corresponding incidence vectors for animal $i$, and $e_{i1}$ is an error term. We did not consider random environmental effects in model (2) but they can be included similarly. Combining equations (1) and (2), and moving the energy sink phenotypes to the left-hand side of the equation, leads to:

$$y_{i1} - \sum_{t=2}^{k} \lambda_{1t} y_{it} = \mu_1 + \mathbf{x}'_{i1}\boldsymbol{\beta}_1 + \mathbf{z}'_{i1}\boldsymbol{a}_1 + e_{i1} \tag{3}$$

Note that, in (3), $y_{i1}$ is a phenotype for DMI, but the fixed and random effects (e.g., $\boldsymbol{\beta}_1$ and $\boldsymbol{a}_1$) pertain to the system phenotype, $y_{i1} - \sum_{t=2}^{k} \lambda_{1t} y_{it}$, which is RFI. This feature contrasts that of a multiple-trait mixed effects model in which the model parameters belong to DMI and energy sinks. Hence, the recursive model can directly estimate RFI genetic values and genetic parameters without taking follow-up partial regression or reparameterization.

Next, each energy sink trait, say $j$, is described similarly by a mixed-effects model:
$$y_{ij} = \mu_j + \mathbf{x}'_{ij}\boldsymbol{\beta}_j + \mathbf{z}'_{ij}\mathbf{a}_j + e_{ij} \tag{4}$$
Stack equation (3) with the mixed-effects models (4) for the energy sink traits, and also take some re-arrangements of incidence matrices and the fixed and random vectors (including the residual vector). Then, we obtain the following recursive structural equation model (RSEM) for individual $i$:
$$\boldsymbol{\Lambda}\mathbf{y}_i = \boldsymbol{\mu} + \mathbf{X}_i\boldsymbol{\beta} + \mathbf{Z}_i\mathbf{a} + \mathbf{e}_i \tag{5}$$
where $\mathbf{y}_i = (y_{i1} \quad y_{i2} \quad \cdots \quad y_{ik})'$, $\boldsymbol{\mu} = (\mu_1 \quad \mu_2 \quad \cdots \quad \mu_k)'$, $\mathbf{e}_i = (e_{i1} \quad e_{i2} \quad \cdots \quad e_{ik})'$, and $\boldsymbol{\beta}$ and $\mathbf{a}$ are vectors of appropriate lengths containing the fixed and random effects, respectively. The vectors of fixed and random effects are re-sorted for each effect by traits within individuals. The incidence vectors are set up accordingly. Let the incidence vectors of fixed effects be the same between traits for each animal. That is, $\mathbf{x}_{i1} = \cdots = \mathbf{x}_{i4} = \mathbf{x}_i$, where $\mathbf{x}_{ij}$ is an incidence vector linking fixed effects to the $j$th phenotype and $\mathbf{x}_i$ is the common incidence vector. Then, $\mathbf{X}_i = \mathbf{x}_i \otimes \mathbf{I}$, where $\mathbf{I}$ is an $k \times k$ identity matrix. Similarly, we have $\mathbf{Z}_i = \mathbf{z}_i \otimes \mathbf{I}$. Finally, the structural matrix is defined as follows:
$$\boldsymbol{\Lambda} = \begin{pmatrix} 1 & -\lambda_{12} & \cdots & -\lambda_{1k} \\ 0 & 1 & \cdots & 0 \\ \vdots & \vdots & \ddots & \vdots \\ 0 & 0 & \cdots & 1 \end{pmatrix} \tag{6}$$
This recursive model belongs to the broad category of recursive and feedback systems for describing the phenotypic relationships between diseases and production in animal breeding (Gianola and Sorensen, 2004; Wu et al., 2007, 2008). Jamrozik et al. (2017) applied recursive modeling to analyze ratio traits, e.g., $y_2/y_1$, where $y_1$ is taken to the baseline trait with an assumed recursive effect on $y_2$. Lu et al. (2015) applied a modified Cholesky decomposition of covariance matrix between DMI and two energy sinks. Intuitively, they showed that the reparameterization implied fully recursive effects from energy sinks and DMI. Neverthelesss, they did not follow the structural equation model but retained a multiple-trait mixed-effects model and obtained partial regression coefficients from estimated covariance matrices. Note that reliably estimating (co)variance components often requires a sufficiently large sample size, regardless of the models. The estimated structural coefficients in RSEM are relatively robust to the estimated variance components because the latter are treated as priors.

The Bayesian implementation of the recursive model for RFI follows Gianola and Sorensen (2004) and Wu et al. (2007, 2010). A simplified algorithm is described below. The RFI phenotypes (i.e., $y_{i1} - \sum_{t=2}^{k} \lambda_{1t} y_{it}$) are uncorrelated with the phenotypes of energy sinks (Kennedy et al., 1993). According to the path theory, a zero phenotypic correlation ($r_p$) between RFI and an energy sink (indexed by j, for $j = 2, \ldots, k$) implies that either 1) $(1 - h_{RFI})(1 - h_j)r_{e_{RFI}e_j} = -h_{RFI}h_j r_{a_{RFI}a_j}$ or 2) $r_{e_{RFI}e_j} = 0$ and $r_{a_{RFI}a_j} = 0$, where $h$ stands for the square root of heritability, and $r_{a_{RFI}a_j}$ and $r_{e_{RFI}e_j}$ are the genetic and residual correlation, respectively, between RFI and energy sink j, assuming a total determination by these two components. The former is a strong assumption which states that genetic and residual correlations between RFI and energy sinks were highly coordinated, and they do not necessarily equal zero. We took the latter approach by forcing the genetic and residual covariance between RFI and energy sinks to be zeros, because we intended to have RFI as a measure of net feed efficiency, independent of energy sinks. That is,
$$\mathbf{G}_0 = \begin{pmatrix} \sigma_{a_1}^2 & 0 & \cdots & 0 \\ 0 & \sigma_{a_2}^2 & \cdots & \sigma_{a_2 a_k} \\ \vdots & \vdots & \ddots & \vdots \\ 0 & \sigma_{a_k a_2} & \cdots & \sigma_{a_k}^2 \end{pmatrix} \tag{7}$$

$$\boldsymbol{R}_0 = \begin{pmatrix} \sigma_{e_1}^2 & 0 & \cdots & 0 \\ 0 & \sigma_{e_2}^2 & \cdots & \sigma_{e_2 e_k} \\ \vdots & \vdots & \ddots & \vdots \\ 0 & \sigma_{e_k e_2} & \cdots & \sigma_{e_k}^2 \end{pmatrix} \tag{8}$$

The covariance matrices between DMI and energy sinks are:

$$\boldsymbol{G}_0^* = \boldsymbol{\Lambda}^{-1} \boldsymbol{G}_0 \boldsymbol{\Lambda}'^{-1}$$
$$= \begin{pmatrix} \sigma_{a_1}^2 + \Delta_a & \lambda_{12}\sigma_{a_2}^2 + \sum_{t \neq 1,2}^{k} \lambda_{1t}\sigma_{a_2 a_t} & \cdots & \lambda_{1k}\sigma_{a_k}^2 + \sum_{t \neq 1,k}^{k} \lambda_{1t}\sigma_{a_k a_t} \\ \lambda_{12}\sigma_{a_2}^2 + \sum_{t \neq 1,2}^{k} \lambda_{1t}\sigma_{a_2 a_t} & \sigma_{a_2}^2 & \cdots & \sigma_{a_2 a_k} \\ \vdots & \vdots & \ddots & \vdots \\ \lambda_{1k}\sigma_{a_k}^2 + \sum_{t \neq 1,k}^{k} \lambda_{1t}\sigma_{a_k a_t} & \sigma_{a_k a_2} & \cdots & \sigma_{a_k}^2 \end{pmatrix} \tag{9}$$

$$\boldsymbol{R}_0^* = \boldsymbol{\Lambda}^{-1} \boldsymbol{R}_0 \boldsymbol{\Lambda}'^{-1}$$
$$= \begin{pmatrix} \sigma_{e_1}^2 + \Delta_e & \lambda_{12}\sigma_{e_2}^2 + \sum_{t \neq 1,2}^{k} \lambda_{1t}\sigma_{e_2 e_t} & \cdots & \lambda_{1k}\sigma_{e_k}^2 + \sum_{t \neq 1,k}^{k} \lambda_{1t}\sigma_{e_k e_t} \\ \lambda_{12}\sigma_{e_2}^2 + \sum_{t \neq 1,2}^{k} \lambda_{1t}\sigma_{e_2 e_t} & \sigma_{e_2}^2 & \cdots & \sigma_{e_2 e_k} \\ \vdots & \vdots & \ddots & \vdots \\ \lambda_{1k}\sigma_{e_k}^2 + \sum_{t \neq 1,k}^{k} \lambda_{1t}\sigma_{e_k e_t} & \sigma_{e_k e_2} & \cdots & \sigma_{e_k}^2 \end{pmatrix} \tag{10}$$

where $\Delta_x = \sum_{t'=2}^{k} \lambda_{1t'}^2 \sigma_{x_{t'}}^2 + \sum_{t'=2}^{k}\left(\lambda_{1t'} \sum_{t \neq 1, t'}^{k} \lambda_{1t}\sigma_{x_t x_{t'}}\right)$, for $x = a$ and $e$, respectively.

The conditional posterior distribution of structural coefficients does not depend on any unknown parameters of energy sinks, assuming zero genetic and residual correlations between RFI and energy sinks. This feature drastically simplifies the posterior inference of structural coefficient matrix and unknown parameters for RFI. Denote $\boldsymbol{\lambda} = (\lambda_{12}, \lambda_{13}, \cdots, \lambda_{1k})'$. We assumed a multivariate normal prior distribution (MVN) for $\boldsymbol{\lambda}$. That is, $\boldsymbol{\lambda}|\lambda_0, \tau^2 \sim MVN(\boldsymbol{1}\lambda_0, \boldsymbol{I}\tau^2)$, where $\boldsymbol{1}$ is a $(k-1) \times 1$ vector of ones, $\boldsymbol{I}$ is a $(k-1) \times (k-1)$ identity matrix, and $\lambda_0$ and $\tau^2$ are hyper-parameters. Then, the conditional posterior distribution of $\boldsymbol{\lambda}$ is also a multivariate normal distribution (Gianola and Sorensen, 2004; Wu et al., 2007), independent of the equations for energy sinks. The conditional posterior means of $\boldsymbol{\lambda}$ are:

$$E(\boldsymbol{\lambda}|else) = \begin{pmatrix} \sum_{i=1}^{n} y_{i2}^2 + \sigma_{e_1}^2 \tau^{-2} & \sum_{i=1}^{n} y_{i2}y_{i3} & \cdots & \sum_{i=1}^{n} y_{i2}y_{ik} \\ \sum_{i=1}^{n} y_{i3}y_{i2} & \sum_{i=1}^{n} y_{i3}^2 + \sigma_{e_1}^2 \tau^{-2} & \cdots & \sum_{i=1}^{n} y_{i3}y_{ik} \\ \vdots & \vdots & \ddots & \vdots \\ \sum_{i=1}^{n} y_{ik}y_{i2} & \sum_{i=1}^{n} y_{ik}y_{i3} & \cdots & \sum_{i=1}^{n} y_{ik}^2 + \sigma_{e_1}^2 \tau^{-2} \end{pmatrix}^{-1}$$
$$\times \begin{pmatrix} \sum_{i=1}^{n} y_{i2}w_{i1} + \sigma_{e_1}^2 \tau^{-2}\lambda_0 \\ \sum_{i=1}^{n} y_{i3}w_{i1} + \sigma_{e_1}^2 \tau^{-2}\lambda_0 \\ \ddots \\ \sum_{i=1}^{n} y_{ik}w_{i1} + \sigma_{e_1}^2 \tau^{-2}\lambda_0 \end{pmatrix} \tag{11}$$

where **else** represents the data and all other unknown model parameters, and $w_{i1} = y_{i1} - (\mu_1 + \boldsymbol{x}'_{i1}\boldsymbol{\beta}_1 + \boldsymbol{z}'_{i1}\boldsymbol{a}_1)$, for $i = 1, \ldots, n$. Similarly, the conditional posterior distribution of location parameters (i.e., fixed and random effects) and scaling parameters (variance components), respectively, for RFI does not involve any unknown parameters for energy sinks either. For example, the conditional posterior means of the RFI location parameters are the following:

$$E\begin{pmatrix}\mu_1\\ \boldsymbol{\beta}_1\\ \boldsymbol{\alpha}_1\end{pmatrix}\bigg|else\bigg) = \begin{pmatrix}\mathbf{1}'\mathbf{y}_1^*\\ \mathbf{x}_1'\mathbf{y}_1^* + \sigma_{e_1}^2\omega^{-2}\boldsymbol{\beta}_0\\ \mathbf{z}_1'\mathbf{y}_1^*\end{pmatrix}\begin{pmatrix}\mathbf{1}'\mathbf{1} & \mathbf{x}_1 & \mathbf{z}_1\\ \mathbf{x}_1' & \mathbf{x}_1'\mathbf{x}_1 + \sigma_{e_1}^2\omega^{-2}\mathbf{I} & \mathbf{x}_1'\mathbf{z}_1\\ \mathbf{z}_1' & \mathbf{z}_1'\mathbf{x}_1 & \mathbf{z}_1'\mathbf{z}_1 + \sigma_{e_1}^2\sigma_{a_1}^{-2}\mathbf{A}^{-1}\end{pmatrix}^{-1} \quad (12)$$

where $\mathbf{y}_1^* = \begin{pmatrix}y_{11} - \sum_{j=2}^k \lambda_{1k}y_{1k}\\ \vdots\\ y_{n1} - \sum_{j=2}^k \lambda_{1k}y_{nk}\end{pmatrix}$, $\mathbf{1}$ is a $n \times 1$ vector of ones, and $\mathbf{I}$ is an identity matrix of appropriate dimensions. In the above, a flat prior is assumed for the overall mean. Multivariate normal prior distributions are assumed for fixed-effect and random-effects: $Pr(\boldsymbol{\beta}_1) = MVN(\boldsymbol{\beta}_0, \omega^2\mathbf{I})$ and $Pr(\boldsymbol{a}_1) = MVN(\mathbf{0}, \mathbf{A}\sigma_{u_1}^2)$, where $\mathbf{A}$ is a numeric additive genetic relationship matrix, $\sigma_{u_1}^2$ is the RFI genetic variance, and $\boldsymbol{\beta}_0$ and $\omega^2$ are hyper-parameters.

To see the link between the recursive model and linear regression for RFI, consider equation (3) and replace the structural coefficients, $\lambda_{1j}$, by partial regression coefficients, $b_j$, for $j = 2, \ldots, k$. If we move all the fixed and random effects to the left-hand side of the equation and keep the energy sinks and the residual on the right-hand side, it becomes:

$$y_{i1} - \mu_1 - \mathbf{x}'_{i1}\boldsymbol{\beta}_1 - \mathbf{z}'_{i1}\boldsymbol{a}_1 = (y_{i2} \quad y_{i3} \quad \cdots \quad y_{ik})\begin{pmatrix}b_2\\ b_3\\ \vdots\\ b_k\end{pmatrix} + e_1 \quad (13)$$

Then, the least-square (LS) solutions of the partial regression coefficients are the following:

$$\begin{pmatrix}\hat{b}_2\\ \hat{b}_3\\ \vdots\\ \hat{b}_k\end{pmatrix}\bigg|\mu_1, \boldsymbol{\beta}_1, \boldsymbol{a}_1\bigg) = \begin{pmatrix}\sum_{i=1}^n y_{i2}^2 & \sum_{i=1}^n y_{i2}y_{i3} & \cdots & \sum_{i=1}^n y_{i2}y_{ik}\\ \sum_{i=1}^n y_{i3}y_{i2} & \sum_{i=1}^n y_{i3}^2 & \cdots & \sum_{i=1}^n y_{i3}y_{ik}\\ \vdots & \vdots & \ddots & \vdots\\ \sum_{i=1}^n y_{ik}y_{i2} & \sum_{i=1}^n y_{ik}y_{i3} & \cdots & \sum_{i=1}^n y_{ik}^2\end{pmatrix}^{-1}\begin{pmatrix}\sum_{i=1}^n y_{i2}w_{i1}\\ \sum_{i=1}^n y_{i3}w_{i1}\\ \vdots\\ \sum_{i=1}^n y_{ik}w_{i1}\end{pmatrix} \quad (14)$$

where $w_{i1} = y_{i1} - \mu_1 - \mathbf{x}'_{i1}\boldsymbol{\beta}_1 - \mathbf{z}'_{i1}\boldsymbol{a}_1$. Note that (11) coincides precisely with (14) if we let $\tau^2 \to \infty$ in (11), which is equivalent to assigning flat priors to structural coefficients in (14). In other words, the conditional posterior means of structural coefficients agree with (or asymptotically equivalent to) the partial regression coefficients in one-step LR, given $\mu_1$, $\boldsymbol{\beta}_1$ and $\boldsymbol{\alpha}_1$, if we ignore the prior values (or the impact of priors diminishes when the data dominates the posteriors). Likewise, the same conclusion holds for the location and scaling parameters between the recursive model and one-step linear regression for RFI.

The heritability for RFI and DMI, respectively, are defined as:

$$h_{RFI}^2 = \frac{\sigma_{a_1}^2}{\sigma_{a_1}^2 + \sigma_{e_1}^2} \quad (15)$$

$$h_{DMI}^2 = \frac{\sigma_{a_1}^2 + \Delta_a}{(\sigma_{a_1}^2 + \Delta_a) + (\sigma_{e_1}^2 + \Delta_{e_r})} \quad (16)$$

The heritability for an energy sink trait takes a similar formula as (15). The genetic correlation between RFI and an energy sink is fixed at zero. The genetic correlation between DMI and an energy sink trait is not zero, which is computed as follows:

$$r_{a_{1t'}} = \frac{\lambda_{1t'}\sigma_{a_{t'}}^2 + \sum_{t \neq 1,t'}^k \lambda_{1t}\sigma_{a_t a_{t'}}}{\sqrt{(\sigma_{a_1}^2 + \Delta_a) \times \sigma_{a_{t'}}^2}}, \text{ for } t' = 2, \ldots, k \quad (17)$$

The data set consisted of 645 first-parity cows with phenotypes, derived from 125 sires and 477 dams and raised in the USDA Beltsville Agricultural Research Center (BARC) Dairy Herd, Beltsville, MD (Connor et al., 2019). The phenotypic data included DMI, MBW, MILKNE, and ΔBW; all obtained as averages over a 42-day trial. Their means (standard deviations) were 28.9 (3.81) kg/d, 113.8 (6.71) $kg^{0.75}$, 21.1 (2.18) Mcal, 0.47 (0.22) kg. Phenotypes were standardized to means of zero and unit variance to facilitate comparing the estimated effects between traits not affected by the units of the traits. The data standardization did not change the phenotypic correlations, which were 0.441 (DMI vs. MBW), 0.556 (DMI vs. MILKNE), 0.166 (DMI vs. ΔBW), 0.132 (MBW vs. MILKNE), 0.193 (MBW vs. ΔBW), and -0.036 (MILKNE vs. ΔBW). We compared two-stage models and one-step models for RFI, implemented by LR and Bayesian RSEM, respectively. LR1 was the stage-one model of the two-stage linear regression, with MBW, MILKNE, and ΔBW as fixed effects. LR2 was a one-step linear regression with three energy sinks and days in milk (DIM: 71, 72, 73, 74, 75, 76, 77) as the fixed effects and individual animal effects as random variables. LR3 had all the model parameters in LR2, plus test weeks (i.e., 143 levels) as an nongenetic random variable. RSEM1, RSEM2, and RSEM were the Bayesian recursive equation models of LR1, LR2, and LR3, respectively, yet with phenotypic recursive effects assumed from energy sinks to DMI. For a Bayesian recursive model, we ran 30 parallel MCMC chains, each consisting of 2,200 iterations, with a burn-in of 2,000 iterations, and thinned every two iterations. We also ran single-trait mixed-effects model analyses on each of these traits, and a multiple-trait, mixed-effects model (MT), with DIM as a fixed effect and test-week and animal effects as random effects.

Markov chain Monte Carlo convergence was examined for the model parameters using the shrink factor (Gelman and Robin, 1992). For each model paramter, let B and W be the between- and within-sequence variance. Then, the Gelman and Rubin's Shrink factor (SF) is computed by:

$$SF = \sqrt{\frac{\frac{L-1}{L}W + \frac{1}{L}B}{W}} \qquad (18)$$

where $L$ is the chain length. The MCMC chains, which were initialized randomly, converged quickly. The shrink factor dropped below 1.1 after 200 iteration, and approached 1.0 after 1,000 iterations (See the graphic abstract). Saved posterior samples after 1,000 iterations were pooled and used to make the posterior inference of unknown parameters.

The estimated effects from energy sinks to DMI (Table 1) agreed very well between LR and recursive models with similar settings (e.g., between LR1 and RSEM1). On standardized phenotypic scales, MILKNE had the largest effects on DMI (0.51 to 0.53), followed by MBW (0.31 to 0.35), and ΔBW had the smallest effect on DMI (0.12 to 0.13). Including different sets of fixed and random effects led to varied RFI definitions, and the estimated partial regression coefficients (or structural coefficients) varied accordingly. Nevertheless, the estimated RFI genetic values agreed very well between a two-stage model and a one-step model. The Spearman's correlation of the estimated RFI genetic values was close to 1 between LR1 and LR3 and between RSEM1 and RSEM3, and re-rankings happened rarely. The Spearman's correlation between LR3 and RSEM3 was 0.998 (Figure 1A). The differences were primarily due to Monte Carlo errors.

The MT model allows for distinguishing between genetic and residual effects. The genetic partial regression coefficients did not agree with the partial regression coefficients (or structural coefficients) obtained from single-trait LR (or recursive models). Nevertheless, the partial regression coefficients estimated from phenotypic (co)variance agreed very well with the partial regression coefficients from LR1 (or the structural coefficients from RSEM1). The multiple-trait, mixed-effects model assumed correlational relationships between the traits, which has no causal interpretation. Nevertheless, a fully-recursiveness system can be assumed based on the modified Cholesky Decomposition (Lu et al., 2015). Consider the phenotypic relationships, for example, in the present example. The $L\Sigma L'$ decomposition implies fully recursive relationships for the traits (ordered by $y_{i4}, y_{i3}, y_{i2},$ and $y_{i1}$). Here, L is the unit lower triangular matrix, which corresponds to the structural coefficient matrix in RSEM, as follows:

$$\Lambda' = L' = \begin{pmatrix} 1 & -b_{12} & -b_{13} & -b_{14} \\ 0 & 1 & -b_{23} & -b_{24} \\ 0 & 0 & 1 & -b_{34} \\ 0 & 0 & 0 & 1 \end{pmatrix} \tag{19}$$

where $b_{jj'}$ is the effect (i.e., partial regression coefficient) from trait j' to trait j, and $y_{ij} = \sum_{j'=1}^{j-1} b_{jj'} y_{ij'}$, for $j = 2, \ldots, 4$. The covariance matrix between the reparameterized variables ($y_{i4}, y_{i3} - b_{34} y_{i4}, y_{i2} - \sum_{j=3}^{4} b_{1j} y_{ij}$, and $y_{i1} - \sum_{j=2}^{4} b_{1j} y_{ij}$) is diagonal, meaning that they are mutually independent. Following the same Bayesian modeling settings to implementing the reparameterized MT model as a Bayesian recursive model, we derive the conditional posterior means of the structural coefficients as follows:

$$E\begin{pmatrix} b_{34} \\ b_{23} \\ b_{24} \\ b_{12} \\ b_{13} \\ b_{14} \end{pmatrix} else = \begin{pmatrix} \frac{\sigma_{e_1}^2}{\sigma_{e_3}^2}\sum y_{i4}^2 + \frac{\sigma_{e_1}^2}{\tau^2} & 0 & 0 & & & \\ 0 & \frac{\sigma_{e_1}^2}{\sigma_{e_2}^2}\sum y_{i3}^2 + \frac{\sigma_{e_1}^2}{\tau^2} & \frac{\sigma_{e_1}^2}{\sigma_{e_2}^2}\sum y_{i3} y_{i4} & & 0 & \\ 0 & \frac{\sigma_{e_1}^2}{\sigma_{e_2}^2}\sum y_{i3} y_{i4} & \frac{\sigma_{e_1}^2}{\sigma_{e_2}^2}\sum y_{i4}^2 + \frac{\sigma_{e_1}^2}{\tau^2} & & & \\ & & & \sum y_{i2}^2 + \frac{\sigma_{e_1}^2}{\tau^2} & \sum y_{i2} y_{i3} & \sum y_{i2} y_{i4} \\ & 0 & & \sum y_{i2} y_{i3} & \sum y_{i3}^2 + \frac{\sigma_{e_1}^2}{\tau^2} & \sum y_{i3} y_{i4} \\ & & & \sum y_{i2} y_{i4} & \sum y_{i3} y_{i4} & \sum y_{i4}^2 + \frac{\sigma_{e_1}^2}{\tau^2} \end{pmatrix}^{-1}$$

$$\times \begin{pmatrix} \frac{\sigma_{e_1}^2}{\sigma_{e_3}^2}\sum y_{i4} w_{i3} + \frac{\sigma_{e_1}^2 \lambda_0}{\tau^{-2}} \\ \frac{\sigma_{e_1}^2}{\sigma_{e_2}^2}\sum y_{i3} w_{i2} + \frac{\sigma_{e_1}^2 \lambda_0}{\tau^{-2}} \\ \frac{\sigma_{e_1}^2}{\sigma_{e_2}^2}\sum y_{i4} w_{i2} + \frac{\sigma_{e_1}^2 \lambda_0}{\tau^{-2}} \\ \sum y_{i2} w_{i1} + \frac{\sigma_{e_1}^2 \lambda_0}{\tau^{-2}} \\ \sum y_{i3} w_{i1} + \frac{\sigma_{e_1}^2 \lambda_0}{\tau^{-2}} \\ \sum y_{i4} w_{i1} + \frac{\sigma_{e_1}^2 \lambda_0}{\tau^{-2}} \end{pmatrix} \tag{20}$$

In the above, the equations for $b_{12}, b_{13}, b_{14}$ in (20) are identical to those for $\lambda_{12}, \lambda_{13}, \lambda_{14}$ in (11), suggesting that both models are equivalent for evaluating RFI, assuming phenotypic recursive effects. However, both models differed in the assumed relationships between energy sinks. The reparameterized MT model is a fully recursive system, as shown in (19). For example, the model by Lu et al. (2015) assumed recursive effects from milk energy (MILKE) on MBW, but the feedback effect from MBW to MILKE did not exist. In contrast, the relationships between energy sinks are correlational in a recursive model.

The estimated partial regression coefficients from the MT model based on the phenothypic covariance matrix coincided precisely with the structural coefficients, showing differences only after the third decimal points, but they can vary depending the data. Based on the standardized phenotypes, the phenotypic partial regressions are derived as follows, which is in a comparable form with (11):

$$\boldsymbol{b} = \boldsymbol{c}_{12}\boldsymbol{V}_{22}^{-1}$$

$$= \left(\sum_{i=1}^{n} y_{i1}y_{i2} \quad \sum_{i=1}^{n} y_{i1}y_{i3} \quad \cdots \quad \sum_{i=1}^{n} y_{i1}y_{ik}\right) \begin{pmatrix} \sum_{i=1}^{n} y_{i2}^2 & \sum_{i=1}^{n} y_{i2}y_{i3} & \cdots & \sum_{i=1}^{n} y_{i2}y_{ik} \\ \sum_{i=1}^{n} y_{i3}y_{i2} & \sum_{i=1}^{n} y_{i3}^2 & \cdots & \sum_{i=1}^{n} y_{i3}y_{ik} \\ \vdots & \vdots & \ddots & \vdots \\ \sum_{i=1}^{n} y_{ik}y_{i2} & \sum_{i=1}^{n} y_{ik}y_{i3} & \cdots & \sum_{i=1}^{n} y_{ik}^2 \end{pmatrix}^{-1} \quad (21)$$

where $\boldsymbol{c}_{12}$ is a vector of the phenotypic covariances between DMI and the energy sink traits, and $\boldsymbol{V}_{22}$ is the phenotypic variance-covariance matrix for the energy sink traits.

Overall, the heritability estimates obtained from RSEM3 and single-trait LR were moderate to high for DMI (0.40-0.49) and MBW (0.59) but low for ΔBW (0.002-0.04) (Table 2). The heritability estimate for RFI was 0.392 by one-step LR, and 0.240 by RSEM3. The heritability estimates for energy sinks were within comparable ranges of previous studies (e.g., Berry and Crowley, 2013; Templeman et al., 2015). RFI heritability estimates were similar to those reported by Connor et al. (2013). They reported an RFI heritability of 0.36 using only the USDA AGIL (Animal Genomics and Improvement Laboratory) data for early lactation cows. Tempelman et al. (2015) reported lower RFI heritability estimates (0.18 ± 0.02) and country-specific estimates ranging from 0.06 up to 0.24. RFI heritability was 0.16 in datasets that included the AGIL data (Lu et al., 2015; Li et al., 2020). Genetic correlations were moderate to high between DMI and RFI, MBW, and MILKNE (0.44 to 0.72), and low between DMI and ΔBW (0.13-0.20) and between MBW and MILKNE (0.15-0.16) (Table 2). The genetic correlations between RFI and energy sinks were forced to be zeros with RSEM3, but they had small values (-0.03-0.01) based on the multiple-trait model.

In conclusion, a recursive model was proposed which postulated recursive effects from energy sinks to DMI, but the feedback or simultaneous effects do not exist. The recursive model is asymptotically to one-step linear regression for RFI, yet extends the analytical capacity to multiple-trait analysis. The present study assumed a single, homogeneous structural coefficient matrix. Model expansion to account for heterogeneous structural coefficient matrices is straightforward, where the conditional distributions for structural coefficients take the same formula but are sampled separately for each subpopulation (Wu et al., 2010; Chitakasempornkul et al., 2020). For modeling RFI alone, the recursive model is equivalent to the reparameterized multiple-trait models based on the modified Cholesky decomposition of (co)variance, but they differed in assumed relationships between energy sinks. The multiple-trait model also allows for distringuishing between genetic and residual relationships. Which assumption is more plausible remains a topic for further investigation. Modeling simultaneous effects between energy sinks and DMI and between energy sinks is also possible, subject to the model identification (Wu et al., 2010).

**Table 1.** Mean or posterior mean (standard deviation or posterior standard deviation) of the estimated effects of three energy sinks on dry matter intake, obtained using different models [1,2,3,4]

| Model | | Effects from energy sinks to dry matter intake | | |
|---|---|---|---|---|
| | | MBW | MILKNE | ΔBW |
| LR | LR1 | 0.351 (0.029) | 0.514 (0.029) | 0.117 (0.030) |
| | LR2 | 0.331 (0.030) | 0.523 (0.029) | 0.123 (0.029) |
| | LR3 | 0.312 (0.029) | 0.534 (0.028) | 0.126 (0.030) |
| RSEM | RESM1 | 0.351 (0.030) | 0.514 (0.029) | 0.117 (0.030) |
| | RSEM2 | 0.331 (0.030) | 0.522 (0.029) | 0.124 (0.029) |
| | RSEM3 | 0.311 (0.030) | 0.530 (0.029) | 0.126 (0.030) |
| MT | Phenotypic | 0.351 | 0.514 | 0.117 |
| | Genetic | 0.257 | 0.746 | 1.185 |

[1] MBW = metabolic body weight; MILKNE = milk net energy; ΔBW = change in body weight.
[2] LR1: linear regression with three energy sinks (MBW, MILKNE, and ΔBW) as the fixed effects; LR2: linear regression with days in milk (DIM) and three energy sinks as fixed effects, and individual animal genetic values as random effects; LR3: LR2 plus test weeks (TW) as a nongenetic random variable.
[3] RSEM1-3: Bayesian recursive models having the same set of model effects as LR1-3.
[4] Phenotypic/Genetic = Partial regression coefficients based on phenotypic and genetic variance-covariance matrices from a multiple-trait model (MT), respectively. DIM were included as fixed-effects, and TW and individual animal genetic values were included as random effects.

**Table 2.** Heritability estimates and genetic correlations for dry matter intake, energy sinks, and residual feed intake, obtained using different models [1,2,3]

| | DMI | MBW | MILKNE | ΔBW | RFI |
|---|---|---|---|---|---|
| | ------ Heritability ------ | | | | |
| ST-LR | 0.489 | 0.592 | 0.355 | 0.044 | 0.392 |
| RSEM3 | 0.400 (0.028) | 0.589 (0.021) | 0.190 (0.025) | 0.002 (0.027) | 0.240 (0.038) |
| | ------ Genetic correlations obtained from RSEM3 (upper) and MT (lower) ------ | | | | |
| DMI | 1 | 0.434 (0.060) | 0.604 (0.055) | 0.129 (0.090) | 0.717 |
| MBW | 0.454 | 1 | 0.145 (0.126) | 0.184 (0.150) | 0 |
| MILKNE | 0.576 | 0.161 | 1 | -0.089 (0.133) | 0 |
| ΔBW | 0.119 | 0.206 | -0.086 | 1 | 0 |
| RFI | 0.713 | -0.022 | 0.012 | -0.033 | 1 |

[1] DMI: dry matter intake; MBW: metabolic body weight; MILKNE: milk net energy; ΔBW: change in body weight; RFI: residual feed intake.
[2] ST-LR (MT): single trait (multiple-trait) mixed-effects model with days in milk as the fixed effects, plus test weeks and individual animal genetic values as the random effects; RSEM3: Bayesian recursive model having days in milk and three energy sinks (MBW, MILKNE, and ΔBW) as the fixed effects, plus test weeks and individual animal genetic values as the random effects.
[3] Posterior standard deviations are shown in brackets.

(A) 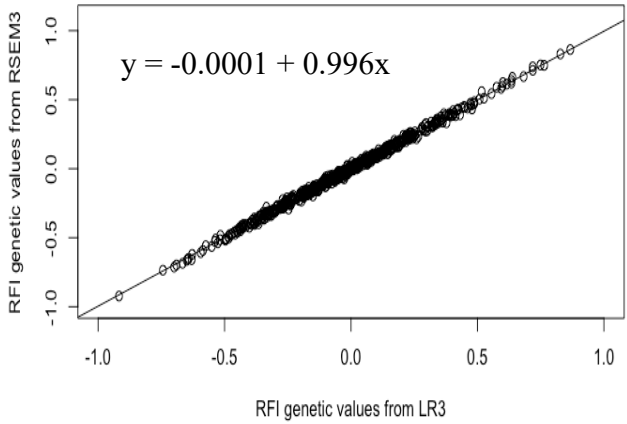   (B) 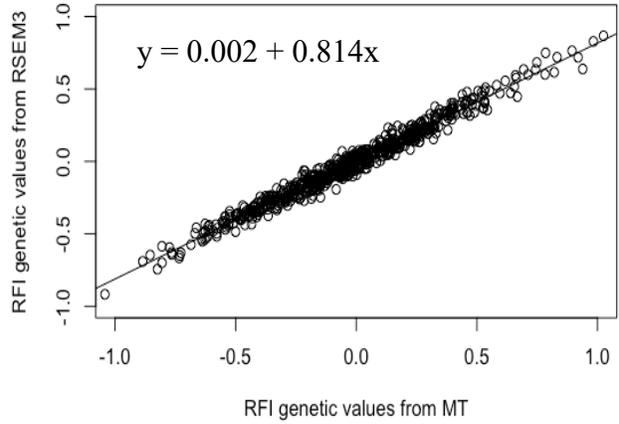

**Figure 1.** Spearman's correlation plots of the estimated genetic animal values obtained from different models: (A) Recursive structural equation model versus one-step linear regression; (B) Recursive model versus multiple-trait mixed-effects model.